# The future of work? Inequality, the advance of Artificial Intelligence, and what can be done about it: A literature review.

Caleb Peppiatt

**Abstract**

Generative Artificial Intelligence constitutes a new wave of automation, parallel to that of robotic automation and computerisation. Broad agreement exists among economists that some job losses will occur, and that humanity is potentially entering into a period of profound change. However, significant uncertainties and disagreements exist concerning a variety of overlapping topics: the share of jobs in which human labour is displaced and/or reinstated through automation; the effects on income inequality; the effects on job satisfaction, and, finally, what policy changes ought to be pursued to reduce potential negative impacts. This literature review seeks to clarify this landscape by mapping out key disagreements between positions, and to identify the critical elements upon which disagreements rest. By surveying the current literature, the effects of AI on the future of work will be clarified.

**I Introduction**

Since the 2010s, breakthrough advances have been achieved in deep learning and transformer architecture. Large Language Models like Chat-GPT have rapidly improved in predictive accuracy following increases in computational investment. The effects of such technological progress are highly contested among economists.

Some writers have concluded that the impact of AI will be utterly transformative ([Gruetzemacher and Whittlestone 2022](#)), and that AI constitutes a "general purpose technology", at least as impactful as the invention of agriculture, electricity, and the Internet, providing the basis for a new Industrial Revolution ([Crafts 2021](#)). Meanwhile, other commentators are more restrained in their predictions, suggesting gains in Total Factor Productivity from AI over the next ten years as being only ~0.71% ([Acemoglu 2024a](#)). Aside from the lack of consensus over the *scale* of impact, substantial disagreement exists over how *positive* such impact might be: optimistic assessments refer to increased productivity and real wages ([Butler et al 2023](#)), or even the rebuilding of the middle class ([Autor Feb 2024](#)). This contrasts, often sharply, with more pessimistic evaluations that describe potential harms to the labour market ([Korinek and Stiglitz 2019](#)), democracy ([Bell and Korinek 2023](#)), income equality and wages ([Bessen et al 2022](#)), developing nations ([Cazzaniga et al 2024](#)), and worker well-being ([Nazareno and Schiff 2021](#)).

Special thanks to Anton Korinek, Philip Trammell, Julian Jacobs and William MacAskill for literature recommendations.

A literature review is particularly helpful when faced with a scholarly landscape with a diverse range of disagreements: keeping track of the particular points of contention is vital for constructive discourse. Often, divergence in thought rests on the key assumptions of models. At other times, it may rest on which empirical data is prioritised. Agreements are also useful to track, as they may indicate surer ground. The review will proceed as follows: In sections II and III, the impact on labour displacement and job exposure to automation will be discussed; section IV will describe channels of increasing inequality which AI may facilitate; section V will look at both the productivity and quality of jobs augmented by AI; section VI will detail both guiding principles that should guide policy-making, and several policies that have been proposed to smooth the effects of AI on work, including the debate over Universal Basic Income; finally, an Appendix will discuss disagreements pertaining to explosive growth models.

## II Theoretical foundations

Economists currently use a task model developed in [Autor, Levy and Murnane 2003](), in which jobs can be atomistically broken down into tasks. Automation is where new technologies enable capital to be substituted for labour in a range of tasks. As described in [Acemoglu and Restrepo 2019](), one can speak of a displacement effect (where capital takes over tasks previously performed by labour), a productivity effect (where there is a more flexible allocation of tasks to capital and labour), and a reinstatement effect (where labour is reinstated into a broad range of new tasks, increasing labour share and demand). If the reinstatement of workers through the creation of new tasks outpaces the displacement of workers, labour demand can rise. Optimistically, one can point towards the fact that the majority of job titles today did not exist in 1940, indicating that new technologies dramatically created new tasks for human labour to perform ([Autor et al 2022]()). Pessimistically, the automation effect has outpaced reinstatement between 1987-2017, leading to a lower demand for labour. Moreover, those workers exposed to displacement due to robots in manufacturing had lower real earnings, and automation accounted for 50-70% of the increase in earnings inequality in those four decades. Regarding productivity effects, AI can increase productivity through performing tasks at lower costs (automation), complementing human labour in tasks, deepening automation further (increasing the productivity of capital), improving research processes, and by creating new tasks for human labour ([Acemoglu 2024a]()). Applying this task model, economists seek to track which tasks can be automated by AI, so as to predict the extent of future displacement and reinstatement effects, and by extension, whether the labour share of the market will increase or decrease ([Autor 2024]()).

## III Job exposure

Tasks can be evaluated as to whether or not they are exposed to automation, where *exposure* is agnostic between *augmentation* (where AI complements human labour, increasing productivity) and *substitution* (where AI replaces human labour for those tasks). Automation technologies operate by substituting capital for labour for a widening array of tasks, and the process of viewing jobs as bundles of tasks has proved fruitful ([Restrepo 2023]()). [Frey and Osborne 2017]() emphasise the routineness of jobs as a key element in exposure to past waves of automation, as robots tend to perform poorly in unstructured environments. They predict that creativity and social intelligence comprise engineering bottlenecks, such that managerial and scientific positions will be safe from AI's advances. Moreover, they show that wages and educational attainment have a strong negative correlation with the probability of displacement, such that low-wage jobs are most at risk. The technological advancements since the publication of this paper cast doubt on its conclusions, as Generative AI can perform a very different set of tasks to previous technologies. OpenAI's recent report, [Eloundou et al 2023](), defines exposure 'as a measure of whether access to a Large Language Model (LLM) would reduce the time required for a human to perform a specific Detailed Work Activity by at least 50%', and concludes that up to 49% of workers could have more than half of their tasks exposed to LLMs. Individuals holding a Bachelor's and/or Master's degree are more exposed, and programming and writing skills are positively correlated with exposure, unlike scientific and critical thinking skills. 19% of workers are in an occupation where over half of its tasks are labelled as exposed, yet as they believe LLMs are General Purpose Technologies that will spawn complementary innovations, this number will likely increase in the coming years. [Felton et al 2023]() employ a similar method of linking AI applications (such as image generation, and translation) to human abilities, and further to jobs as a weighted combination of the human abilities, in order to create an AI Occupational Exposure measure. They find a strong positive correlation between their exposure score and mean wages in an occupation, such that higher wage jobs are more likely to be exposed, including industries like education, accounting, software publishing, and legal services. The European Commission's Technical Report, [Tolan et al 2021](), map generic tasks from worker surveys to 14 cognitive abilities, and these abilities to a list of 328 AI-related benchmarks that indicate progress in machine learning. The results again showed that high-skilled occupations such as medical doctors and teachers are more exposed, and that intellectual abilities are more exposed than social ones, a finding which others independently predicted ([Deming 2017]()). However, the importance of social skills may not be sustained in the long-run, if Generative AI eclipses humans in this area, being free from cognitive biases. [Webb 2020]()'s methodology instead compared patents (supplying information about what technologies do) with text descriptions of jobs, finding that low-skill jobs are most exposed to robots, middle-skill jobs are most exposed to software, and upper-middle-skill jobs are most exposed to AI, such as clinical laboratory technicians and chemical engineers. [Eloundou et al 2023](), [Felton et al 2023](), [Tolan et al 2021](), and [Webb 2020]() each, despite using different methods of comparing the task contents of jobs to AI capabilities, find that higher-skill jobs, and higher-middle income jobs, associated with writing skills, are most exposed to AI's advances, a striking difference to [Frey and Osborne 2017]() that indicates how the tasks exposed to Generative AI are highly different from those exposed to previous waves of automation.

Three significant caveats apply to this general conclusion. Before concluding that tasks exposed to AI will be automated, one has to first consider the cost-effectiveness of doing so. [Svanberg et al 2024]() find that only 23% of worker compensation exposed to AI computer vision would be cost-effective for firms to automate due to large upfront costs, and even if computer vision costs halved annually, the effects of AI automation would be smaller than existing job destruction effects. Secondly, although an impressive range of tasks are exposed to automation, few jobs can be *fully* automated using AI, such that full automation may be less significant than the reorganisation of tasks ([Brynjolfsson et al 2018]()). Third, exposure is ambiguous between tasks in which workers are complemented, and tasks in which workers are displaced, two processes which have significantly different impacts on income equality.

**IV The effects of AI's advances on inequality**

The impact on income inequality partly depends on the results of section III, and on the degree of automation that will transpire. If higher-wage jobs are automated rather than merely augmented, high-wage workers would be displaced, such that inequality could potentially be reduced as a measure of the difference between the 90th and 10th percentiles of incomes. Under Michael Webbs' model, wages might decline 8-14%. On the other hand, the inequality at the top of the income distribution would increase, as measured as a ratio between the 99th and the 90th percentile, as capital-owners would get wealthier ([Webb 2020](#)). If lower-wage workers are instead displaced (which does not seem to be the case from advances in Generative AI), this could further increase inequality: historically, as routine tasks in industries experienced rapid robotic automation over the past four decades, US wages for these workers dropped ([Acemoglu and Restrepo 2022](#)). [Bessen et al 2022](#) found that workers in Dutch firms that decided to automate lost around 9% of annual wage earnings after 5 years, a real wage loss of around 3500 euros, affecting older, middle-income workers the most. Meanwhile, if higher-wage jobs are more complementary to AI augmentation, as found in [Cazzaniga et al 2024](#), income disparity could be further increased.

For explanatory purposes, we can thus create a four-part taxonomy: i) if upper-middle to high-income workers are displaced by AI, income inequality between these workers and lower-wage workers might decrease; ii) if such upper-middle income workers are instead augmented by AI, increasing their productivity (as shown in section V.I), then income inequality would increase; iii) if lower-income workers are displaced by automation (as shown historically by robotics), then income inequality will increase as labour share decreases; and iv) if lower-income workers are instead augmented by AI, this could increase their productivity, and thereby reduce income-inequality. From section III, it seems that upper-middle income workers are most exposed to AI, and section V.I. will show that for many such workers, productivity will increase, potentially leading to a further disparity between incomes.

Two further channels of inequality are commented upon in the literature: Firstly the "great divergence" between developing and developed nations may further increase as a result of the adoption of, and investment in, AI. This discussion also depends on the degree of job exposure and complementarity among jobs. [Cazzaniga et al 2024](#) finds that around 40% of workers worldwide are in high-exposure occupations, whilst this increases to 60% of jobs in advanced countries, suggesting that advanced economies are more susceptible to labour market shifts from AI adoption. Their AI Preparedness Index ranks countries based on their digital infrastructure, innovation and economic integration, human capital and labour market policies, and regulation and ethics, finding that wealthier economies are generally better prepared to adopt AI, such that AI adoption could potentially widen the digital divide and global income disparity. On a simpler level, only a tiny number of companies possess frontier AI labs, and so would enjoy the greatest productivity booms. [Alonso et al 2020](#) argue that GDP per capita in developing nations can decrease due to advances in AI, if, as would plausibly occur, capital flows uphill from developing countries to advanced economies to finance accumulated robotic capital.

Secondly, the economic inequality resulting from AI can damage democracies. [Bell and Korinek 2023](#) describe the erosion of democracy and increasing inequality as a vicious cycle: high inequality sows discontent and populism, and wealthy individuals can capture policy agenda, whilst democratic processes can mitigate inequality through mechanisms of representation and accountability. Highly concentrated wealth translates to political power and undermines political equality, a point thoroughly made in [Christiano 2012](#). Aside from potentially massive windfall profits generated by leading AI companies, such companies may gain increasing power through accumulating data of consumer preferences ([Acemoglu 2022](#)). [Boix 2022](#) suggests several policies to counteract dangers to democracy, including strong antitrust policies to preempt the formation of closed elites, the limiting of campaign donations by corporations, and the opening up of borders to facilitate a South-North migration flow and thereby counteract increased interregional inequalities.

**V Job productivity and worker well-being**

**V.I.** How does integrating AI into a workplace augment workers? [Brynjolfsson et al's 2023](#) focused case study on customer support agents show that an AI conversational assistant improved productivity of workers by an average of 14%, including, most notably, a 34% improvement for low-skilled workers. Similar productivity effects are found in [Noy and Zhang's 2023](#) experiment of 453 professionals completing writing tasks using help from ChatGPT: participants who used the LLM took less time and their work was deemed superior in quality by their peers. Moreover, the variance in the quality of written output decreased. Participants using ChatGPT also reported that they enjoyed completing their task more, however, this could be due to the novelty of interacting with it for the first time. [Peng et al 2023](#) extend these general findings of increased productivity to programmers using the GitHub Copilot. Microsoft's annual report, [Butler et al 2023](#), also details productivity gains in a variety of working environments, although experiments are still in their infancy. However, one should not conclude that augmentation increases productivity in all tasks, as it is highly difficult to tell - for both researcher, employer and worker - which tasks are complementary. A complex business case in which workers used GPT-4, for instance, was much less successful than a control group, yet such workers were confident in their results ([Dell'Acqua et al 2023](#)). [Autor Feb 2024](#) believes that such augmentation, by decreasing the variance in productivity between workers, might empower a larger set of workers to gain expertise and perform high-stakes decision-making, thereby rebuilding a middle class that had been damaged by computerisation.

**V.II.** Productivity itself is a poor proxy for worker well-being, which has been comparatively under-researched. Technological complementarity, such as the integration of AI into workflows, can impact worker well-being through altering the content of work (which tasks are performed), the context of work (such as the time and space in which work is performed), and the psychological contract between employers and workers. [Nazareno and Schiff 2021](#) use job satisfaction, stress, health, and job security as key measures for worker well-being, and found that the integration of AI into workflows did not produce a positive, symbiotic environment in which worker freedom and creativity increased. Worker stress was generally reduced, yet job satisfaction declined, perhaps indicating the rise of easy yet repetitive tasks. The stress reduction effect of automation has weakened since 2002, with increases in stress perhaps being explained by increased surveillance and control over workers. Effects on health and job security were far less significant. The view that worker-AI complementarity can reduce both worker autonomy and the subjective meaning assigned to tasks has further qualitative support in [Stephanie Bell's 2023](#) surveys of sub-Saharan data annotators that switched to using ML software, and of US factory workers who switched to automated warehouses. Tasks became easier and simpler, and by extension became less engaging. Such surveys, however, also reveal general increases in the speed, accuracy, efficiency and productivity of workplaces that used AI, with many Indian call-centre workers appreciating the advice and time-saving benefits offered by AI call-assistants.

One major point of agreement among commentators is that the impact of AI on the wellbeing of workers is not deterministic, and instead depends on managerial and executive decision-making. Negative outcomes associated with workplace AI, such as pressure from constant monitoring, overly aggressive targets and shift-scheduling, are not inevitabilities. Bell highlights the importance of a company's business model: designing roles to be performed with little training, autonomy, and average tenure is a natural fit with surveillance technology and high worker stress; workers should instead be empowered early on in the implementation process with roles that retain worker input. [Acemoglu and Johnson 2024](#) likewise stresses that increased productivity is not necessarily good for workers, and uses the Industrial Revolution as a case study to emphasise that the balance of power between workers and employers is both crucial and malleable.

**VI Policy proposals**

**VI.I.** Which guiding principles and considerations should influence the debate on policies? Whilst profound disagreement exists over which policies are empirically justified, commentators generally agree that policies ought to be pro-work; that they ought to reduce inequality; and that many policies have to exist in tandem.

Firstly, policy-makers ought to gain a holistic understanding of the utility of work, and to consider the implications of job displacement on displaced workers themselves. Involuntary job loss generally results in a variety of economic and non-economic outcomes ([Brand 2015](#)). Economically, this tends to include a long period of unemployment, a cumulative lifetime earnings loss of roughly 20%, and transitions into lower quality jobs with increased job instability. Non-economic effects include increased lower social involvement in wider communities, higher levels of depressive symptoms, worse self-reported health, lower self-esteem and educational attainment of the workers' children, and increased chance of hospitalisation, divorce, and suicide. Moreover, work itself carries a variety of non-monetary benefits, including the structuring of time, social contact, individual self-esteem, and a stronger locus of control ([Dijkema and Gunderson 2019](#)). From this data, many contend that policies should be pro-work, working to reduce involuntary job loss; at the same time, work should no longer be viewed as grounding an individual's status, as job loss regularly leads to worsened social perception and self-esteem. In an age of increasing automation, policies should recognise the vital importance of work as a source of meaning, structured activity, and social interaction, and so should encourage work, rather than merely plan for its potential abolition (McAfee and Brynjolfsson 2016). Given the importance of work, technology should be steered to complement human labour, to allow workers to perform their duties more efficiently, and to invent new human-focused tasks performed by labour (Korinek and Stiglitz 2020). One way to increase the demand for unskilled labour heavily affected by automation are through investments in infrastructure, such as public transportation systems, or digital infrastructure, which are additionally important in reducing the digital divide between nations (Korinek and Stiglitz 2021). At the same time, the right public narrative surrounding work should be cultivated, such that the negative social perceptions and psychological effects of job loss are reduced (Korinek and Juelfs 2022).

Policies should seek to reduce channels of inequality, a notion shared by all writers covered in section IV, albeit for diverse reasons. Korinek and Stiglitz 2019 are concerned with the increasing concentration of surplus earned by innovators, alongside rising unemployment levels, and reduced welfare. Bell and Korinek 2023 and Boix 2022 motivate policies to reduce inequality by surveying damaging effects on democracy. Robeyns 2023's limitarian thesis, a unique form of inequality reduction, is motivated both by concerns for democracy as well as the redistributive potential for meeting unmet urgent needs. O'Keefe et al 2020's proposed "Windfall Clause", a voluntary commitment for frontier AI firms to share profits that are unprecedentedly large, is motivated by job losses, the potential formation of global oligopoly, and the provision of a financial buffer that would smooth the transition to advanced AI. The IMF's reports and policy recommendations (Cazzaniga et al 2024 and Alonso et al 2020) largely seek to address disadvantages to developing nations. Beraja and Zorzi 2022 find that the government should seek to slow down automation whilst workers reallocate, even if one simply wishes to make automation *more* efficient and alleviate borrowing frictions. Indeed, Korinek and Stiglitz 2019 show that 'compensating workers for losses imposed by technological progress is a question of economic efficiency,' such that redistribution programmes make the economy's allocation more, rather than less, efficient. Acemoglu and Lensman 2024 find that automation should be slowed down and new technologies adopted gradually as an optimal strategy for minimising expected damages on the labour market. The sheer diversity of independent reasons for reducing the inequality brought on by advances in AI confirm its importance as a policy priority.

Thirdly, policies ought to extend across various sectors, with no silver bullet solution existing. A diverse range of harms resulting from advances in AI have been proposed (Acemoglu 2022), and so a broad range of policies working in tandem to target specific issues is necessary. Several regulatory reforms have been proposed: In terms of property laws, shortening the terms of patent protection for IPs allows society to enjoy the benefits of new technologies, at lower prices, sooner (Korinek and Stiglitz 2019). Privacy regulations can prevent some data misuses, tort law can ensure AI companies are legally liable for faults, and antitrust law can prevent oligopolies forming as AI companies grow in wealth and influence (Agrawal et al 2019). The optimal balance between too much regulation (discouraging innovation), and too little (failing to prevent adverse effects) is unlikely to be reached in the short-term, such that flexibility rather than rigidity in policy-making is vital (McAfee and Brynjolfsson 2016).

**VI.II.** Three major policies that aim to reduce inequality are frequently debated on empirical grounds: guaranteed incomes, minimum wages, and the taxation of capital.

One major source of contention is over the efficacy of Universal Basic Income (UBI) as a solution to rising inequality and job insecurity. UBI is meant to guarantee a minimal level of support without disincentivising work. One can distinguish between *general* objections to UBI, which apply more broadly to the concept, and *specific* objections to UBI, which target it as an inadequate response to the harms of AI. The general objection that UBI reduces incentives to work has not stood up to empirical scrutiny in the long-term unconditional transfer schemes in Kenya ([Banerjee et al 2023](#)). The Kenyan experiments did not find negative outcomes, although this is perhaps a small and unrepresentative sample. Another common objection, that UBI would decrease efficiency by increasing equity, has not been substantiated by European guaranteed living income experiments ([Pressman 2005](#)). In terms of benefits, UBI also has been said to provide financial assistance to women, enabling them to leave abusive relationships ([Schulz 2017](#)), and possibly counterbalances strains on familial relationships brought on by involuntary joss losses ([Brand 2015](#)). Guaranteed income also likely impacts the psychological contract between workers and employers, such that UBI may make it easier for employees to leave a job if they are dissatisfied ([Perkins et al 2021](#)), a positive outcome for ensuring worker wellbeing in times of technological change ([Acemoglu and Johnson 2024](#)). Stronger critiques of UBI have been raised by writers who directly relate it to a discussion on AI-caused labour market changes. For a given amount of money to be used on redistribution, UBI likely shifts resources away from the very poor (as all recipients are equally targeted) compared to current poverty-reduction strategies ([Goolsbee 2019](#)). This is not a fatal objection, however, as proponents of UBI rarely seek to entirely replace current social welfare strategies, and recognise that UBI would be insufficient as a sole welfare solution. A less tractable problem with UBI, and with many other policies that seek to reduce inequality, is its universality: the challenge of making it global in nature, so as to reduce international inequality ([Korinek and Juelfs 2022](#)). Proposals like the Windfall Clause ([O'Keefe et al 2020](#)) may, if AI companies become signatories, and if the clause was triggered, finance such a universal scheme, yet the administrative practicalities remain highly speculative. Finally, [Przegalinska and Wright 2021](#) argue that UBI's focus on employment and unilateral transfers is too narrowly focused, such that policy-makers should consider subsistence activities, proprietorship and financial investment as well. Even if jobs disappear, individuals will have time to increase the parts of their real income that stem from these sources.

A second contentious policy recommendation that seeks to address inequality is raising minimum wages. The general problem with this policy is that when labour becomes more expensive, companies tend to use less of it ([McAfee and Brynjolfsson 2016](#)). An analysis of the published literature on minimum wages in the United States indicates that higher minimum wages do have negative employment effects ([Neumark and Shirley 2022](#)), although the evidence for low-wage industries is less clear. A third policy seeks to tax capital, given that the capital share of the economy will likely increase with advances in AI. If one taxes a fixed factor of production, as long as the supply elasticity of capital is sufficiently low, this can enable redistribution with minimal market distortions ([Korinek and Stiglitz 2019](#); [Agrawal et al 2019](#)). [Guerreiro et al 2020](#) find that taxing firms for replacing a human worker with a robot reduces the non-routine wage premium, thereby helping redistribute income toward routine workers who are most affected by labour-displacement. [Seamans 2021](#) strongly disagrees with a robot tax, arguing that there is little evidence that robots are directly substituting for human labour, and that the term "robot" is critically ambiguous, and can lead to disproportionate burdens on some industries rather than others. Seamans instead suggests a broader tax raise on capital investments, as the tax code currently taxes labour more harshly, favouring automation

([Acemoglu et al 2020](#)). An alternative adjustment to capital would be the establishment of national sovereign wealth funds: firms would be asked by governments to contribute to a fund which is owned by everyone in a nation ([Korinek and Stiglitz 2021](#)). Such wealth funds increase the net worth of a state, which is a valuable proxy measure for what we pass on to future generations, and so can be a vehicle for achieving intergenerational equity (Atkinson 2015). OpenAI CEO Sam Altman also proposes an "American Equity Fund", wherein every citizen is an owner of company shares, an effective tax on capital that aligns incentives between companies, investors, and citizens ([Altman 2021](#)). A major point of failure for such equity funds is that they are national in nature, and the value of such AI companies are highly concentrated in particular nations, with the majority of frontier AI companies existing on American soil. If successful, such an equity fund would dramatically increase the wealth of US citizens, but would further increase inequality between nations.

**VII Conclusion**

Several major points of agreement and of contention have been identified in this literature review. First, section III showed that several studies with independent methods agree that higher-middle income, and higher-middle skill occupations, associated with intellectual skills, are most exposed to advances in Generative Artificial Intelligence. Second, section V.I show that such jobs will mostly be complemented by AI, with productivity boosts observed in various experiments in 2023-4. Third, section IV showed that, although there is disagreement as to how complementary exposure to AI is likely to be, there are various ways in which income inequality may increase. Fourth, section VI.I showed that some agreement exists over guiding principles surrounding policy-making: namely, that policies ought to be pro-work, that they should seek to reduce inequality, and that there is no "silver bullet" solution, necessitating multiple policy angles. Section VI.II, meanwhile, showed that three common policy recommendations - UBI, minimum wages, and capital taxation and ownership - are each heavily debated on empirical grounds. As such, it is perhaps unwise for policy-makers to place all their eggs in the basket of one of these policies. Finally, the appendix overviewed some of the purported productivity effects of AI, finding that although scenarios of explosive growth are theoretically achievable, more research needs to be done as to the fundamental limits of research-productivity and labour accumulation before their plausibility can be assessed.

**Appendix - Discussion of Explosive Growth**

There are massive disagreements among economists as to the productivity effects of AI's advances. A discussion of productivity gains is necessary for considering inequality, as even if income inequality is high, real incomes could theoretically be large enough that this is not seen as a significant social problem. [Acemoglu 2024a](#) suggests a very modest forecast of an increase in Total Factor Production of 0.55% in the next ten years for tasks exposed to AI, and an increase in GDP in the range of 0.9-1.1%. However, as [Maxwell Tabarrok](#) has noted, this model is overly conservative: it relies on current technological data (GPT-4), is restricted to a decade's timespan, and neglects the effects of deepening automation. [Abel et al. 2021](#) suggests that generative AI could add 2.6-4.4 trillion USD across the 63 cases they analysed, primarily through saving time on knowledge work. Also considering a short-term time-frame of ten years, Goldman Sachs 2023 suggests that global GDP could rise by 7% as an upper figure, although this relies upon extensive new job creation and significant technological advances. Polling of economists on whether growth rates of real per capita income in the US will increase due to AI reveals a substantial minority of experts optimistic, yet a larger minority was highly uncertain (especially in the qualitative comments). Berg et al 2018 find that under a broad range of views concerning productivity gains, automation is very good in the long-run for growth, and very bad for equality. Due to a strong positive feedback loop between robot and non-robot capital accumulation, real per capita income could increase between 30-240% in the long-run. However, the larger the increase in GDP, the less equitable its distribution, as the labour share of the economy decreases.

Contrasting with short-term forecasts and estimates, a minority of growth economists suggest that advances in AI could lead to explosive, superexponential economic growth. A related notion is that of the singularity, which, whilst having an abundance of meanings (Sandberg 2013), refers in the economic literature to growth rates increasing without bound, yet remaining finite at any one point in time (Aghion et al 2017). This contrasts with the standard picture in growth economics, which claims that growth rates will stagnate due to population growth rates declining.

There are two major processes by which a growth explosion is meant to be achieved. The first is that AI can automate the discovery of new ideas: ideas are non-rivalrous, and can be subject to increasing returns. If AI could augment or replace people as researchers, then the research growth rate as an economic input could experience a growth explosion (Jones 2022). There is experimental evidence that LLMs are highly productive at idea generation, with GPT-4 generating innovative ideas at a faster and cheaper rate than students at a top university (Girota et al 2023). Moreover, AI can augment R&D by increasing capital investment in research: in fact, given certain assumptions, a profit maximising firm would allocate 29-44% of its total R&D budget to computational capital, greatly increasing economic growth (Besiroglu et al 2024). The central question in evaluating this idea-based argument concerns the explanation for why research productivity has fallen. Importantly, Bloom et al 2020 have shown that research productivity has been declining, such that the number of researchers required to generate new ideas has risen each decade. They infer that ideas simply get harder to find - that there are diminishing returns in acquiring ideas, perhaps due to an increasing "burden of knowledge" that researchers have to learn before covering new ground (Clancy 2021), or perhaps because most of the easy ideas - low-hanging fruit - have already been found. If there is some fundamental characteristic about ideas that means finding more of them is limited, then AI may not facilitate acquiring ideas with increasing returns, such that a growth explosion will not be reached. It might even be the case that AI is unhelpful in improving R&D, due to restrictive regulation so as to avoid dangerous ideas being uncovered, or due to the share of mundane knowledge work increasing, or due to a reproducibility crisis (Almeida et al 2024). Ultimately, the idea-based argument for a growth explosion is theoretically sound, yet there is high uncertainty surrounding the limits of idea-acquisition and research productivity.

The second process by which AI could fuel a productivity explosion is if human labour was replaced by digital workers. Labour then becomes accumulable, resulting in increasing returns to scale in production (Davidson 2021). An analysis of the long-term gross world product indicates that superexponential growth is not unthinkable, as gross world product is relatively unstable (Roodman 2020). The extent to which digital workers will replace human labour might be hindered by regulatory legislation restricting access to data and computational resources ([Erdil and Besiroglu 2023](#)). Another obstacle might be Baumol's "cost disease," which predicts that sectors with rapid productivity growth would see their share of GDP decline ([Aghion et al 2017](#)). A further major objection to the substitution of production is that there is very little evidence that we are headed for an economic singularity, in any economic sector ([Nordhaus 2021](#)): if we are to project trends, it would be over a century before superexponential growth is reached. There are, however, a variety of sound models by which a growth explosion may be achieved through labour becoming accumulable, with a diverse set of extreme consequences ([Korinek and Trammell 2023](#)). If digital workers *were* to replace human workers, as the labour share decreases and capital accumulates, the income share would asymptotically vanish. For the labour share to remain positive in the long-run, either it must be impossible for some sectors to be fully automated, or it must be impossible for robots to self-replicate, or else there must be non-homothetic demand that progressively favours sectors where humans are harder to replace ([Ray and Mookherjee 2022](#)). Scenarios in which AI are fully capable of replacing human workers tend to be extreme in their conclusions, with real wages potentially plummeting, tending towards zero ([Korinek and Stiglitz 2019](#), [Korinek and Juelfs 2022](#), [Korinek and Suh 2024](#)). As such, there needs to be more research done on the critical disagreements between economists on this topic: specifically, on the conditions under which digital workers could replace human workers, and the point at which this would lead to superexponential growth.

---